\documentclass[notitlepage,showpacs,showkeys]{revtex4-1}
\usepackage{amsfonts}
\usepackage{amssymb}
\usepackage{amsmath}
\usepackage{graphicx}
\usepackage{float}
\usepackage{subcaption}
\usepackage[font={footnotesize,it}]{caption}

\begin{document}

\preprint{}
\title{Magnetic Morris-Thorne wormhole in 2+1-dimensions }
\author{S. Habib Mazharimousavi}
\email{habib.mazhari@emu.edu.tr}
\author{Zahra Amirabi}
\email{zahra.amirabi@emu.edu.tr}
\author{M. Halilsoy}
\email{mustafa.halilsoy@emu.edu.tr}
\affiliation{Department of Physics, Eastern Mediterranean University, Gazima\u{g}usa,
North Cyprus, Mersin 10 - Turkey.}

\begin{abstract}
In the context of $2+1-$dimensional gravity coupled to a particular
nonlinear electrodynamics (NED), we obtain a class of traversable /
Morris-Thorne type wormhole solutions. The problem is reduced to a single
function dependence in which the shape function acts as generator to the
wormholes. The field ansatz is pure magnetic and the nonlinear Lagrangian is 
$\sqrt{F_{\mu \nu }F^{\mu \nu }}$ i.e. the square root of the Maxwell
Lagrangian. In $2+1-$dimensions the source-free pure magnetic non-linear
Maxwell equation with square-root Lagrangian is trivially satisfied. The
exotic energy density is found explicitly and the flare-out conditions are
emphasized.
\end{abstract}

\maketitle

\section{Introduction}

Although the topic of spacetime wormholes was popularized in modern times by
the seminal works of Morris and Thorne and Visser \cite{MT}, the original
idea traces back to the 'bridge' of Einstein and Rosen \cite{ER} and even to
the embedding diagram of Flamm \cite{Flamm}. In brief, it is a hypothetical
shortcut spacetime tunnel that connects vastly distant points belonging
either to the same or different universes. Classical physics prohibits such
travels due to instability unless an exotic matter source is taken for
granted to support the tunnel. Similar to black holes wormholes are also
exact solutions to Einstein's field equations. The idea of time travel
through a wormhole, however, transcends classical considerations. To draw a
rough analogy we may refer to the Art of Escher \cite{Escher}, where in the
same picture birds transmute into fishes etc. While this transmutation takes
place in our minds, for the sake of Art, physical theory of wormholes
demands far more than this kind of visualization. In brief an observer can
travel from one universe into the other through a traversable wormhole can
also connect distant parts of the same universe. Yet in this analogy we can
say that in the realm of wormholes Einstein meets Escher. Wormholes demand
physical transition between vastly separated points in warped spacetime in
which curvature of spacetime plays the principal role in Einstein's
relativity and given the suitable energy-momentum such a travel becomes
possible according to the laws of physics. Another interesting development
took place recently in connection with wormholes: the Einstein-Rosen (ER)
bridge and the spooky interaction of quantum particles known as
Einstein-Podolski and Rosen (EPR) pair may be related. Symbolically this
situation has been summarized by $ER=EPR$ \cite{EPR}, which may serve to
connect wormholes with the realm of quantum theory.

For these reasons we took wormhole physics seriously and attempted to
construct these objects on physical, i.e., non-exotic matter \cite{MH1}. To
certain extent we obtained results that employ non-circular / non-spherical
throat topology in the wormholes \cite{MH2}. In addition to the energy
matters recently we have also revised the well-known flare-out conditions 
\cite{MH3}.

In this paper we resort to the non-linear electromagnetism to provide a
possible source for our traversable wormhole in $2+1-$dimensions. This is
the square-root Lagrangian of the Maxwell invariant which breaks the scale
invariance. Being a square-root expression our electromagnetic field is
automatically pure magnetic, i.e., we have $F_{r\theta }\neq 0,$ as the only
non-zero electromagnetic field component. The energy density turns out to be
exotic and under this condition we present exact wormhole solutions. In \cite%
{MT} the idea of a traversable wormhole is introduced and the flare-out
conditions which every traversable wormhole must satisfy are also found. In
accordance with \cite{MT} the general line element of a traversable,
circularly symmetric wormhole in $2+1-$dimensions is written as%
\begin{equation}
ds^{2}=-e^{2\Phi \left( r\right) }dt^{2}+\frac{dr^{2}}{1-\frac{b\left(
r\right) }{r}}+r^{2}d\theta ^{2},
\end{equation}%
in which $\Phi \left( r\right) $ is called the redshift function and $%
b\left( r\right) $ stands for the shape function of the wormhole. If we
consider the location of the throat which connects two distant spacetimes,
at $r=b_{0}$, the flare-out conditions state that: i) $b\left( r_{0}\right)
=r_{0}$ and ii) $b^{\prime }\left( r\right) <\frac{b\left( r\right) }{r},$
where prime means $\frac{d}{dr},$for $r\geq r_{0}.$ Although, in \cite{MT} a
throat is a gate between two asymptotically flat spacetimes (i.e., $%
\lim_{r\rightarrow \infty }\Phi =0$ and $\lim_{r\rightarrow \infty }\frac{%
b\left( r\right) }{r}=0$) this condition is not necessary due to the
existence of non-asymptotically flat spacetimes as the solutions of the
Einstein's gravity coupled to different matter fields such as dilaton \cite%
{2,3,4,5}. Therefore, the only constraint on $\Phi $ is to be finite in the
domain of $r_{0}\leq r.$ Having the Einstein's equations and flare-out
conditions all satisfied yields a negative energy density $\rho <0.$ This
can be seen from the $tt$ component of the Einstein's equation where $%
G_{t}^{t}=T_{t}^{t}$ ($8\pi G=c=1$)$.$ From the line element (1) and the
fact that $T_{t}^{t}=-\rho ,$ one finds%
\begin{equation}
\frac{\left( b^{\prime }-\frac{b}{r}\right) }{2r^{3}}=\rho .
\end{equation}%
From (2) one easily observes that with the flare-out condition fulfilled
i.e., $b^{\prime }-\frac{b}{r}<0,$ the energy density becomes negative.
Hence, the traversable wormholes are supported by exotic matter which
violates the null energy condition.

Wormholes in $2+1-$dimensions, relatively, received less attention than the $%
3+1-$dimensional counterparts \cite{6,7,Rah,8,9,10,11,12,13}. It worths to
mention that the first work on $2+1-$dimensional wormholes was studied by
Perry and Mann in \cite{7}.

In this paper we consider traversable wormholes in $2+1-$dimensions
supported by a nonlinear electrodynamic (NED) matter source. The nonlinear
Maxwell's Lagrangian which is employed in this study, namely the square root
of the Maxwell Lagrangian is of the form given in \cite{14} which was
developed further in \cite{15,16,17,18,19,20,21,22,23,24}.

\section{Morris-Thorne type wormhole in $2+1-$dimensions}

\begin{figure}[h]
\includegraphics[width=90mm,scale=0.7]{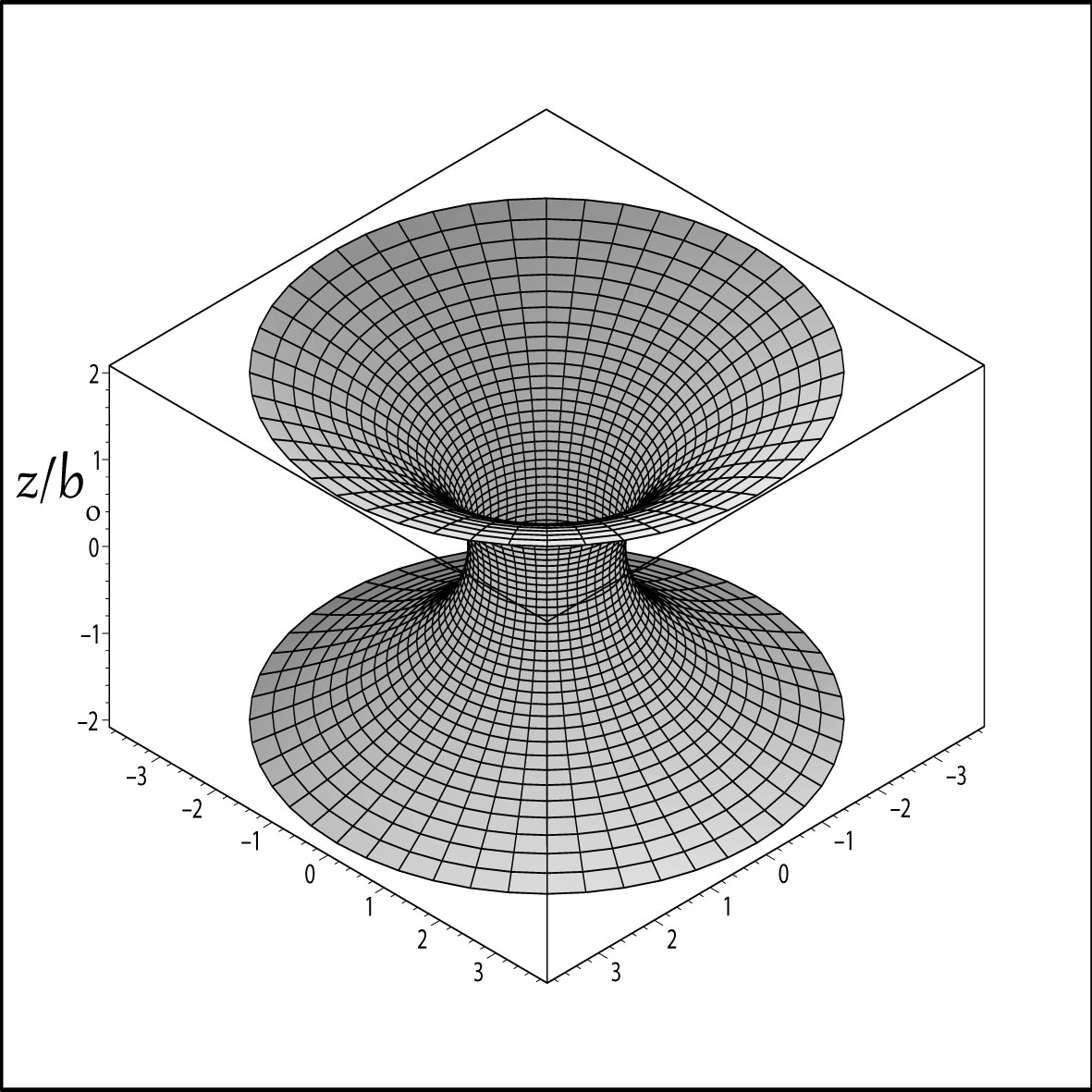}
\caption{$\frac{z}{b_{0}}$ versus $\frac{r}{b_{0}}$ and $\protect\theta $ in
cylindrical coordinates (See Eq. (8)). We note that at $r=b_{0}$, $z=0$ is
where the throat lies. At the location of the throat the magnitude of the
curvature scalar i.e. $\left\vert R\right\vert $ is maximum while at large $%
r $ it goes to zero. The negative energy density gets its maximum value also
at the throat while at large $r$ it vanishes.}
\end{figure}

Let's start with the line element%
\begin{equation}
ds^{2}=-dt^{2}+dl^{2}+\left( l^{2}+b_{0}^{2}\right) d\theta ^{2}
\end{equation}%
which we wish to call the Morris-Thorne type wormhole (MTtW) in $2+1-$%
dimensions. Note that in 3+1-dimensions such a wormhole was introduced by
Ellis \cite{Ellis}. Herein, $b_{0}$ is a real parameter, $-\infty <t<\infty
, $ $-\infty <l<\infty $ and $0\leq \theta \leq 2\pi .$ The Ricci scalar of
MTtW%
\begin{equation}
R=-\frac{2b_{0}^{2}}{\left( l^{2}+b_{0}^{2}\right) ^{2}}
\end{equation}%
is clearly negative and the geometry is regular everywhere. The Ricci scalar
admits an absolute / relative minimum located at $l=0$ while for $%
l\rightarrow \pm \infty $ it vanishes. Furthermore, the only nonzero
component of the Einstein's tensor is given by%
\begin{equation}
G_{t}^{t}=\frac{b_{0}^{2}}{\left( l^{2}+b_{0}^{2}\right) ^{2}}
\end{equation}%
which yields%
\begin{equation}
\rho =-\frac{b_{0}^{2}}{\left( l^{2}+b_{0}^{2}\right) ^{2}}=\frac{R}{2}
\end{equation}%
where $\rho $ is the energy density of the matter, supporting the MTtW. It
can easily be seen that $\rho $ behaves the same as $R$ such that a minimum
occurs at $l=0.$ Upon taking a time slice of the spacetime (3) and embedding
the result in cylindrical coordinates as%
\begin{equation}
ds^{2}=dl^{2}+\left( l^{2}+b_{0}^{2}\right) d\theta
^{2}=dr^{2}+dz^{2}+r^{2}d\theta ^{2}
\end{equation}%
yields $l^{2}+b_{0}^{2}=r^{2}$ and $\left( \frac{dz}{dr}\right) ^{2}=\frac{%
b_{0}^{2}}{r^{2}-b_{0}^{2}}.$ These clearly imply that $r^{2}\geq b_{0}^{2}$
and 
\begin{equation}
z=\pm \int_{b_{0}}^{r}\frac{dx}{\sqrt{\frac{x^{2}}{b_{0}^{2}}-1}}=\pm
b_{0}\ln \left( \frac{r}{b_{0}}+\sqrt{\frac{r^{2}}{b_{0}^{2}}-1}\right) ,
\end{equation}%
which is the same paraboloid of revolution as in the $3+1-$dimensional MTtW 
\cite{MT}. We note that $r^{2}=b_{0}^{2}$ is equivalent to $l^{2}=0.$ In
Fig. 1 we plot $\frac{z}{b_{0}}$ in terms of $\frac{r\text{ }}{b_{0}}$ and $%
\theta .$ This figure supports the idea of having a throat located at $z=0$
corresponding to $r=b_{0}$ and therefore $l=0,$ where $R_{\min }=2\rho
_{\min }=-\frac{2}{b_{0}^{2}}.$ To complete this section we add that a
transformation of the form we introduced above, i.e., $%
l^{2}+b_{0}^{2}=r^{2}, $ helps us to find the more familiar form of the line
element of the MTtW as%
\begin{equation}
ds^{2}=-dt^{2}+\frac{dr^{2}}{1-\frac{b_{0}^{2}}{r^{2}}}+r^{2}d\theta ^{2}
\end{equation}%
which suggests that $\Phi =0$ and $b\left( r\right) =\frac{b_{0}^{2}}{r}$ in
(1). Once more we stress that in (9), $b_{0}\leq r$ so that $r=b_{0}$
corresponds to $l=0$ which defines the location of the throat. There is no
need to state also that the flare-out conditions are perfectly satisfied.

\section{MTtW in non-linear electrodynamics (NED) coupled to gravity}

Let's start with the line element of a static and circularly symmetric
spacetime given in (1) in which $\Phi \left( r\right) $ and $b\left(
r\right) $ depend only on $r$. We note that $-\infty <t<\infty ,$ $0\leq
r<\infty $ and $\theta \in \left[ 0,2\pi \right] .$ The action for gravity
coupled to NED in $2+1-$dimensions is given by ($8\pi G=1=c$) 
\begin{equation}
S=\frac{1}{2}\int d^{3}x\sqrt{-g}\left( R-2\Lambda +\mathcal{L}\right)
\end{equation}%
in which $R$ is the Ricci scalar, $\Lambda $ the cosmological constant, and $%
\mathcal{L}=\alpha \sqrt{F}$ stands for the nonlinear Maxwell's Lagrangian.
Note that $\alpha $ is a coupling constant and $F=F_{\mu \nu }F^{\mu \nu }$
with $F_{\mu \nu }=\partial _{\mu }A_{\nu }-\partial _{\nu }A_{\mu }$ is the
Maxwell's invariant. Let us remind that historically it was Born and Infeld 
\cite{BI} who considered a non-linear version of electromagnetic Lagrangian
that survived to present time. In their approach one could obtain the linear
Maxwell Lagrangian as a limiting procedure. In our choice of square-root
Maxwell Lagrangian, however, we shall have no such a limit. Once we let $%
\alpha \rightarrow 0$ with $\Lambda \neq 0$ we arrive at the BTZ \cite{BTZ}
black hole solution. for the choice $\Lambda =0$ (with $\alpha =0$) in
2+1-dimensions we recover nothing but the flat spacetime. It should also be
added that the original motivation of NED was to eliminate the divergences
in electromagnetic field due to the point charges. We comment that recently
Einstein's gravity coupled minimally to the nonlinear Maxwell's Lagrangian
of the form $\mathcal{L}\left( F\right) \sim F^{k}$, received attentions
from different aspects \cite{14, 15,16,17,18,19,20,21,22,23,24}. Here, we
consider $k=\frac{1}{2}$ i.e., $\mathcal{L}\left( F\right) \sim \sqrt{F}$
with a pure magnetic field. Let us add that the particular power $k=\frac{3}{%
4}$ corresponds to the scale invariant case, i.e. invariance under $x_{\mu
}\rightarrow \lambda x_{\mu }$ and $A_{\mu }\rightarrow \frac{1}{\lambda }%
A_{\mu },$ for $\lambda =$constant, in $2+1-$dimensions. Our choice $k=\frac{%
1}{2}$ therefore breaks the scale invariance with physical consequences. It
should also be remarked that $\mathcal{L}\left( F\right) \sim \sqrt{F}$ in
flat spacetime had been studied long ago by Nielsen and Olesen \cite{25} in
string theory while 't Hooft \cite{tHooft} highlighted a linear potential
term to be effective toward confinement. Our choice of the Maxwell's 2-form
is just a magnetic field of the form%
\begin{equation}
\mathbf{F=}B\left( r\right) dr\wedge d\theta
\end{equation}%
in which $B\left( r\right) $ is a function of $r$ to be found. Breaking the
scale invariance the Lagrangian $\sqrt{F}$ has the interesting property that
it confines geodesics \cite{Guendelman}. The source-free nonlinear-Maxwell's
equation for the specific Lagrangian chosen, is given by%
\begin{equation}
d\left( \frac{^{\ast }\mathbf{F}}{\sqrt{F}}\right) =0
\end{equation}%
in which%
\begin{equation}
F=2B^{2}\frac{\left( 1-\frac{b\left( r\right) }{r}\right) }{r^{2}}
\end{equation}%
with its dual 1-form%
\begin{equation}
^{\ast }\mathbf{F=}\frac{Be^{\Phi }\sqrt{1-\frac{b\left( r\right) }{r}}}{r}%
dt.
\end{equation}%
Hence, (12) yields%
\begin{equation}
e^{\Phi }=const.
\end{equation}%
and consequently%
\begin{equation}
\Phi =C
\end{equation}%
in which $C$ is an integration constant. We note that, $e^{C}$ can be easily
absorbed in time $t$ and therefore without loss of generality we set $C=0.$
Next, the Einstein's equations with a cosmological constant $\Lambda $ are
given by%
\begin{equation}
G_{\mu }^{\nu }+\frac{1}{3}\Lambda \delta _{\mu }^{\nu }=T_{\mu }^{\nu }
\end{equation}%
in which 
\begin{equation}
T_{\mu }^{\nu }=\frac{\alpha }{2}\left( \mathcal{L}\delta _{\mu }^{\nu }-4%
\mathcal{L}_{F}F_{\mu \lambda }F^{\nu \lambda }\right) .
\end{equation}%
Upon (11) and (18), one finds $T_{r}^{r}=T_{\theta }^{\theta }=0$ and 
\begin{equation}
T_{t}^{t}=\frac{\alpha }{2}\sqrt{F}=\frac{\alpha }{\sqrt{2}}B\frac{\sqrt{1-%
\frac{b\left( r\right) }{r}}}{r}.
\end{equation}%
Furthermore, with $\Phi =0,$ the only nonzero component of the Einstein's
tensor is 
\begin{equation}
G_{t}^{t}=-\frac{\left( b^{\prime }-\frac{b}{r}\right) }{2r^{2}},
\end{equation}%
in which a prime stands for the derivative with respect to $r.$ We obtain as
a result, $\Lambda =0$ in order to have the $rr$ and $\theta \theta $
components of the Einstein-Maxwell's equations satisfied. Next, we consider
the $tt$ component of the Einstein-Maxwell's equation which reads 
\begin{equation}
-\frac{\left( b^{\prime }\left( r\right) -\frac{b\left( r\right) }{r}\right) 
}{2r^{3}}=\frac{\alpha }{\sqrt{2}}B\left( r\right) \frac{\sqrt{1-\frac{%
b\left( r\right) }{r}}}{r}.
\end{equation}%
This equation gives a relation between the magnetic field $B\left( r\right) $
and the shape function $b\left( r\right) .$ In other words, a general class
of solutions is determined by (21) such that the redshift function is zero
while the shape function and the magnetic field satisfy the constraint 
\begin{equation}
B\left( r\right) =-\frac{\sqrt{2}}{2\alpha }\frac{\left( rb^{\prime
}-b\right) }{r^{3}\sqrt{1-\frac{b}{r}}}.
\end{equation}%
From this expression one finds that a possible throat is located at $r=b_{0}$%
. We note that (22) may or may not result in a wormhole. For instance $%
b\left( r\right) =0$ yields $B\left( r\right) =0$ and the spacetime becomes
flat. Hence, to have a traversable wormhole one should find specific
function for $b\left( r\right) $ such that the Morris-Thorne's flare-out
conditions are satisfied. In the following sections we give two specific
wormhole solutions.

\subsection{MTtW}

In the first example we consider the shape function to be of the form $%
b\left( r\right) =\frac{b_{0}^{2}}{r}$ in which $b_{0}$ is the location of
the throat. This shape function has been found above in MTtW in $2+1-$%
dimensions (see Eq. (9)). Having $b\left( r\right) ,$ one finds the form of
the magnetic field which is determined as%
\begin{equation}
B\left( r\right) =\frac{b_{0}^{2}\sqrt{2}}{\alpha r^{3}\sqrt{r^{2}-b_{0}^{2}}%
}.
\end{equation}%
This is a singular function of $r$ such that at the location of the throat
it diverges. The Maxwell invariant, however, $F=\frac{4b_{0}}{r^{10}}$ is
finite at $r=b_{0}.$ In addition to that, at large $r$ the magnetic field
vanishes to give an asymptotically flat limit for the wormhole.

The Ricci and Kritchmann scalars, respectively, are 
\begin{equation}
R=-\frac{2b_{0}^{2}}{r^{4}}
\end{equation}%
and%
\begin{equation}
K=\frac{4b_{0}^{4}}{r^{8}}
\end{equation}%
which are clearly regular at the throat. We wish to proceed now with the
investigation of geodesic completeness \cite{41} in order to verify that
divergence of the magnetic field at the throat is of no significance. The
geodesics Lagrangian is 
\begin{equation}
L=-\frac{1}{2}\dot{t}^{2}+\frac{1}{2}\left( 1-\frac{b_{0}^{2}}{r^{2}}\right)
^{-1}\dot{r}^{2}+\frac{1}{2}r^{2}\dot{\theta}^{2}
\end{equation}%
where a dot stands for derivative with respect to the proper time. The first
integrals of $t$ and $\theta $ equations are 
\begin{eqnarray}
\dot{t} &=&E=cons. \\
r^{2}\dot{\theta} &=&\ell =cons..  \notag
\end{eqnarray}%
The timelike geodesics ($L=-\frac{1}{2}$)give%
\begin{equation}
\frac{dr}{dt}=\frac{1}{E}\sqrt{\left( E^{2}-1-\frac{\ell ^{2}}{r^{2}}\right)
\left( 1-\frac{b_{0}^{2}}{r^{2}}\right) }.
\end{equation}%
The radial geodesics ($\ell =0$) yields the hyperbolic curve 
\begin{equation}
r\left( t\right) =\sqrt{b_{0}^{2}+\alpha _{0}^{2}t^{2}}
\end{equation}%
in which $\alpha _{0}^{2}=1-\frac{1}{E^{2}}.$ It is observed that for $%
-\infty <t<\infty ,$ we have $b_{0}\leq r<\infty $ which reflects the
completeness of geodesics in the wormhole spacetime.

Next, for $\ell \neq 0$ we arrive at the expression%
\begin{equation}
\int\nolimits_{r_{0}}^{r}\frac{rdr}{\sqrt{\left( \left( E^{2}-1\right)
r^{2}-\ell ^{2}\right) \left( 1-\frac{b_{0}^{2}}{r^{2}}\right) }}=t-t_{0}
\end{equation}%
with the initial time constant $t_{0}.$ This can be reduced to an elliptic
integral form and naturally the geodesic completeness is valid here as well.

Finally, the tidal forces at the throat can be analyzed through the
geodesics deviation equation%
\begin{equation}
\frac{D^{2}\xi ^{i}}{d\tau ^{2}}=-R_{jkl}^{i}\xi ^{k}\frac{dx^{j}}{d\tau }%
\frac{dx^{\ell }}{d\tau }
\end{equation}%
where $\xi ^{i}$, ($i=1,2$) are displacements along the radial and angular
directions. One obtains%
\begin{equation}
\frac{D^{2}\xi ^{1}}{d\tau ^{2}}=\frac{b_{0}^{2}\ell ^{2}}{r^{8}}\xi ^{1}
\end{equation}%
and%
\begin{equation}
\frac{D^{2}\xi ^{2}}{d\tau ^{2}}=\frac{b_{0}^{2}}{4}\xi ^{2}\left( E^{2}-1-%
\frac{\ell ^{2}}{r^{2}}\right) \left( 1-\frac{b_{0}^{2}}{r^{2}}\right)
\end{equation}%
which indicate the finiteness of tidal forces in the vicinity of the
wormhole throat.

\subsection{Generalized MTtW}

As a second example we consider the shape function to be of the form $%
b\left( r\right) =\frac{b_{0}^{\mu +1}}{r^{\mu }}$ in which $\mu $ is a
free, real parameter. Note that in order to have the flare-out conditions
satisfied we must impose $\mu >-1.$ The case $\mu =1$ has already been
considered in our example A. Among other possibilities, we consider $\mu =0$
which yields $b\left( r\right) =b_{0}.$ Consequently, the magnetic field and
the energy density become%
\begin{equation}
B\left( r\right) =\frac{b_{0}\sqrt{2}}{2\alpha r^{3}\sqrt{1-\frac{b_{0}}{r}}}
\end{equation}%
and 
\begin{equation}
\rho =-\frac{b_{0}}{2r^{3}}.
\end{equation}%
It should be stressed here also that the Maxwell invariant in the present
case is $F_{\mu \nu }F^{\mu \nu }=\frac{b_{0}}{\alpha ^{2}r^{8}},$ which is
regular at the throat. One finds that the proper distance takes the form%
\begin{equation}
l\left( r\right) =\pm \left[ r^{2}\sqrt{1-\frac{b_{0}}{r}}+\frac{b_{0}}{2}%
\ln \left( \frac{2r}{b_{0}}\left( 1+\sqrt{1-\frac{b_{0}}{r}}\right)
-1\right) \right]
\end{equation}%
and the shape function is%
\begin{equation}
z\left( r\right) =\pm 2\sqrt{b_{0}\left( r-b_{0}\right) }.
\end{equation}%
The Ricci and Kretchmann scalars become now%
\begin{equation}
R=-\frac{\left( \mu +1\right) b_{0}^{\mu +1}}{r^{\mu +3}}
\end{equation}%
and%
\begin{equation}
K=\frac{\left( \mu +1\right) ^{2}b_{0}^{2\left( \mu +1\right) }}{r^{2\left(
\mu +3\right) }}
\end{equation}%
which imply that the throat is a regular hypersurfacet. Given the analysis
of the previous section it is not difficult to anticipate that the tidal
forces / accelerations are finite in this generalized MTtW model as well.
Due to the power $\mu $ however the integrals of geodesics will not be any
simpler.

\section{Conclusion}

We constructed a class of traversable wormhole solutions in the theory of
gravity coupled to nonlinear electrodynamics in $2+1-$dimensions. A similar
model of wormhole with an anisotropic fluid source was considered in \cite%
{Culetu}. For specific choice of the shape function the solution is the
Morris-Thorne type wormhole in $2+1-$dimensions which shares most of its
properties with its $3+1-$dimensional version. The matter source which
supports our wormhole solution is a pure magnetic field of the form given in
(22). The square-root of pure magnetic Maxwell Lagrangian provides automatic
satisfaction of the non-linear Maxwell equation in $2+1-$dimensions.
Confining of geodesics is another interesting property of such a square-root
Lagrangian \cite{Guendelman}. We comment that the magnetic field diverges at
the throat and vanishes fast with $r\rightarrow \infty .$ A particularly
simple example with $b\left( r\right) =b_{0}=const.$ is considered. In this
example also the magnetic field diverges at the throat while the Maxwell
invariant $F_{\mu \nu }F^{\mu \nu }$ is finite at the throat. Next, we
consider a more general ansatz which involves an arbitrary parameter $\mu .$
We note that divergence of the magnetic field at the throat was used in \cite%
{Lobo} as a counter argument against existence of such 2+1-dimensional
wormholes. The only singularity of the problem lies at $r=0$ which is a
naked spacetime singularity but since the wormhole condition stipulates that 
$r\geq b_{0}$ the singularity at $r=0$ where the scalar curvature invariants
diverge remains ineffective for particle geodesics. Finally, we should add
that the class of solutions found in this paper consists of a large number
of solutions which only depends on the form of $b\left( r\right) $. Any
choice of $b\left( r\right) $ satisfying the flare-out conditions acts as a
generator and gives rise to a new Morris-Thorne type wormhole. The fact that
we work in the reduced $2+1-$dimensions simplifies the problem to a great
extend.\ In $3+1-$dimensions obviously wormholes can't be generated from a
single throat function.

\end{document}